\documentclass{article}
\usepackage{epsfig}
\usepackage{amsfonts}

\begin{document}

\begin{center}

{\huge{\bf The main frequencies of solar core natural
oscillations}}

\bigskip
\author "B.V.Vasiliev
\bigskip

Institute in Physical-Technical Problems, 141980, Dubna, Russia
\bigskip

{sventa@kafa.crimea.ua}
\end{center}

\bigskip


\bigskip

\begin{abstract}
The oscillations of a spherical body consisting of hot
electron-nuclear plasma are considered. It is shown that there are
two basic modes of oscillations. The estimation of the main
frequencies of the solar core oscillation gives a satisfactory fit
of the calculated spectrum and the measurement data.
\end{abstract}
\bigskip

\section{Introduction. The main parameters of star cores}

The substance of a star interior exists as hot dense
electron-nuclear plasma. In the general case, the equilibrium of a hot
plasma in a gravitational field can be written as
\begin{equation}
{\bf \nabla} P + \gamma {\bf g} + \rho {\bf E}= 0\label{gen}
\end{equation}
where ${\bf \nabla} P$ is the pressure gradient, $\gamma$ is the mass
density  and $\rho$ is the electric density induced by the gravity in
plasma, satisfying
\begin{equation}
4\pi\rho=div~{\mathbf{E}},\label{divE}
\end{equation}
\begin{equation}
-4\pi G \gamma=div~\mathbf{g}.\label{divg}
\end{equation}
It is conventionally accepted to think that the equilibrium exists
at ${\bf E}=0$ and
\begin{equation}
{\bf \nabla} P + \gamma {\bf g}= 0.
\end{equation}
At that the gradient pressure induces the increasing of the
density and the temperature depthward a star. According to the
virial theorem the full energy of  star is equal to one half of
its gravitational (potential) energy. It means the order of value
of the full energy
\begin{equation}
{\mathbb{E}}_{({\bf {E}=0})}\approx -\frac{GM_{star}^2}{R_{star}},
\end{equation}
where $M_{star}$ and $R_{star}$ are the mass and the radius of a star.
However, it is possible to see that Eq.({\ref{gen}}) can be
reduced to another equilibrium condition
\begin{equation}
\gamma {\bf g} + \rho {\bf E}= 0
\end{equation}
at ${\bf \nabla}P=0$.  Because of a high plasma density into a
star core \cite{1}, the significant part of a star substance is
concentred in its core, and  the star energy in the order of value
is
\begin{equation}
{\mathbb{E}}_{({\bf \nabla} P=0)}\approx
-\frac{GM_{star}^2}{R_{core}}.
\end{equation}
As a core radius $R_{core}<R_{star}$, the state with ${\nabla
P=0}$ is energetically preferable for the central core of a plasma
body. The problem of equilibrium plasma in self-gravitationing
field is considered in detail in \cite{1}.

The absence of the pressure gradient results in constant plasma
density. It is not difficult to see, that the plasma state with
constant density corresponds to the energy minimum. Indeed, taking
into account of inter-particle interaction, the free energy of the
hot plasma is \cite{2}
\begin{equation}
{\mathbb{F}}_{plasma} =F_{id}+\frac{\pi^{3/2}}{4}\biggl(\frac{r_B
e^2}{kT}\biggr)^{3/2}N_e n_e - \frac{2\pi
^{1/2}}{3}\biggl(\frac{\langle Z\rangle+1}{kT}\biggr)^{3/2}e^3 N_e
n_e^{1/2},
\end{equation}
where $F_{id}$ is the free energy of the ideal gas, $N_e$ is the
full number of electrons in a system, $n_e$ is the electron
density. In this equation the second item describes the correction
for quantum properties of electron gas and third item describes
the electron-ion interaction in plasma. The full energy of ideal
gas doesn't depend on particle density by definition and the
derivative from $F_{id}$ is negligible small. At a constant full
number of particles in the system and at a constant temperature,
the equilibrium state exists at the minimum of energy

\begin{equation}
\biggl(\frac{\partial {\mathbb{F}}_{plasma}}{\partial
n_e}\biggr)_{N,T}=0,\label{dedn}
\end{equation}
which allows one to obtain the steady-state value of the density of hot
non-relativistic plasma

\begin{equation}
{\Theta}=\frac{16(\langle Z\rangle+1)^{3}}{9 \pi^2 r_B^3}\simeq
2\cdot 10^{24}(\langle Z\rangle+1)^3 ~ cm^{-3}, \label{neq}
\end{equation}
where $r_B=\frac{\hbar^2}{me^2}$ is the Bohr radius, $\langle
Z\rangle$ is the averaged charge of nuclei plasma is composed from.

The equilibrium conditions give possibility to calculate all main
parameters of the star core \cite{1}:

the equilibrium temperature:

\begin{equation}
{\mathbb{T}}=\biggl(\frac{10}{\pi^4}\biggr)^{1/3}(\langle
Z\rangle+1)\frac{\hbar c}{kr_B}\approx 2  \cdot 10^7(\langle
Z\rangle+1)~K;\label{4a}
\end{equation}

the equilibrium mass of core:
\begin{equation}
{\mathbb{M}}=1.5^6\biggl(\frac{10}{\pi^{3}}\biggr)^{1/2}
\biggl(\frac{\hbar c}
{Gm_p^2}\biggr)^{3/2} \biggl\langle \frac{Z}{A}\biggr\rangle^2 m_p\nonumber\\
\approx 6.47~ M_{Ch}~
\biggl\langle\frac{Z}{A}\biggr\rangle^2\label{M},
\end{equation}
where $M_{Ch}=\biggl(\frac{\hbar c}{G
m_p^2}\biggr)^{3/2}m_p=3.42\cdot 10^{33} g$ is the Chandrasechar
mass, $\langle A\rangle$ is averaged mass number of nuclei
 plasma consists from, $m_p$ is the proton mass;

the equilibrium radius of the star core:

\begin{eqnarray}
{\mathbb{R}}= \frac{(3/2)^3}{2}
\biggl(\frac{10}{\pi}\biggr)^{1/6}\biggl(\frac{\hbar c}{G
m_p^2}\biggr)^{1/2} \frac{r_B}{(\langle Z\rangle+1){\langle
A/Z\rangle}}.\label{s10}
\end{eqnarray}

and  the pressure into a core:
\begin{equation}
{\mathbb{P}} =
\frac{G{\mathbb{M}}^2}{8\pi{\mathbb{R}}^4}.\label{pr}
\end{equation}

It is important to underline that the steady-state parameters of a
core are depending only on chemical composition of a star, that is
expressed through two variable parameters $\langle Z\rangle$ and
$\langle A/Z\rangle$. These parameters are unknown apriori.

\section{The sound speed in a hot plasma}
The pressure of a high temperature plasma is a sum of the plasma
pressure (ideal gas pressure) and the pressure of black radiation:
\begin{equation}
P=n_e kT +\frac{a}{4} (kT)^4,\label{pr2}
\end{equation}
and its entropy is
\begin{equation}
s=\frac{1}{m'}~ln~\frac{(kT)^{3/2}}{n_e} +\frac{a(k
T)^3}{n_e},\label{s}
\end{equation}
where $m'=\frac{A}{Z}m_p$ is the mass related to one electron and
$a={4\pi^2}/{45\hbar^3 c^3}$.

The sound speed $c_s$ can be expressed by a Jacobian \cite{3}:
\begin{equation}
c_s^2=\frac{D(p,s)}{D(\rho,s)}=\frac{\biggl(\frac{D(p,s)}{D(n_e,T)}\biggr)}{\biggl(\frac{D(\rho,s)}{D(n_e,T)}\biggr)}
\end{equation}
or
\begin{equation}
c_s=\biggl\{\frac{5}{3}\frac{k{T}}{\langle A/Z\rangle m_p}
\biggl[1+\frac{2a^2(kT)^6}{5n_e[n_e+2a(kT)^3]}\biggr]\biggr\}^{1/2}
\end{equation}

For  $T={\mathbb{T}}$ and $n_e=\Theta$ we have:
\begin{equation}
\frac{a(k{T})^3}{n_e}=\frac{a(k{\mathbb{T}})^3}{\Theta}=\frac{1}{2}~.
\end{equation}
 Finally we obtain:
\begin{equation}
c_s=\biggl\{\frac{5}{3}\biggl(\frac{10}{\pi^4}\biggr)^{1/3}\frac{(\langle
Z\rangle+1)\hbar c}{\langle A/Z\rangle m_p
r_B}[1.05]\biggr\}^{1/2} \approx 5.42~10^7~\biggl(\frac{\langle Z
\rangle +1}{\langle A/Z \rangle }\biggr)^{1/2}~ cm/s~.\label{cs}
\end{equation}

\section{The basic elastic oscillation of a spherical core}
Star cores consist of a dense high temperature plasma which is a
compressible matter. The basic mode of elastic vibrations of a
spherical core is related with its radius oscillation. For the
description of this type of oscillation, the potential $\phi$ of
displacement velocities $v_r=\frac{\partial \psi}{\partial r}$ can
be introduced  and the motion equation can be reduced to the wave
equation expressed through $\phi$ \cite{3}:
\begin{equation}
c_s^2\Delta\phi=\ddot\phi,
\end{equation}
and a spherical derivative for periodical in time oscillations
$(\sim e^{-i\Omega_s t})$
 is:
\begin{equation}
\Delta\phi=\frac{1}{r^2}\frac{\partial}{\partial
r}\biggl(r^2\frac{\partial \phi}{\partial
r}\biggr)=-\frac{\Omega_s^2}{c_s^2}\phi~.
\end{equation}
It has the finite solution for the full core volume including its
center
\begin{equation}
\phi=\frac{A}{r}sin \frac{\Omega_s r}{c_s},
\end{equation}
where $A$ is a constant. For  small oscillations, when
displacements on the surface $u_R$ are small $(u_R/R=v_R/ \Omega_s
R\rightarrow 0)$ we obtain the equation:
\begin{equation}
tg \frac{\Omega_s {\mathbb{R}}}{c_s}=\frac{\Omega_s
{\mathbb{R}}}{c_s}
\end{equation}
which has the solution:
\begin{equation}
\frac{\Omega_s {\mathbb{R}}}{c_s}\approx4.49.
\end{equation}
Taking into account  Eq.({\ref{cs}})), the main frequency
of the core radial elastic oscillation is
\begin{equation}
\Omega_s =
4.49\biggl\{\frac{10.5}{(3/2)^7\pi}\biggl[\frac{Gm_p}{r_B^3}\biggr]\biggl\langle\frac{A}{Z}\biggr\rangle
\biggl(\langle Z\rangle +1\biggr)^3\biggr\}^{1/2}.\label{qb}
\end{equation}
It can be seen that this frequency depends on $\langle{Z}\rangle$
and $\langle{A}/{Z}\rangle$ only.

Some values of  frequencies of radial sound oscillations ${\cal
F}=\Omega_s/2\pi$  calculated from this equation for selected
$A/Z$ at $Z=1$ and $Z=2$ are shown in third column of Table 1.

\bigskip
{\hspace{10cm} Table 1.}

\begin{tabular}{||c|c|c||c|c||}\hline\hline
&&${\cal F},mHz$&&${\cal F},mHz$\\
Z&A/Z&&star&\\
& &(calculation Eq.({\ref{qb}}))&&measured\\ \hline
1&1&0.78&$\eta~Bootis$&0.85\\ \hline
&&& The Procion$(A\alpha~CMi)$&1.04\\
1&2&1.10& &|| \\
&&&$\beta~Hydri$&1.08\\ \hline 2&2&2.02&&\\\hline
&2.5&2.25&&\\
2&&&$\alpha~Cen~A$&2.37\\\
&3&2.47& & ||\\\hline
2&3.5&2.67&&\\
2&4&2.85&&\\
2&4.5&3.02&&\\\hline 2&5&3.19&The Sun&3.23\\ \hline\hline
\end{tabular}

\bigskip

The measured frequencies of surface vibrations for some of stars
\cite{4} are shown in right part of this table. The data for $\nu
~Indus$ and $\xi~Hydrae$  also exists \cite{4}, but characteristic
frequencies of these stars are below 0.3 mHz and they have some
another mechanism of excitation, probably. One can conclude from
the data of Table 1 that the core of the Sun is basically composed
by helium-10 . It is not a confusing conclusion, because according
to Eq.({\ref{pr}}), the pressure which exists inside the solar
core amounts to $10^{19} {dyne}/{sm^2}$ and it is capable to
induce the neutronization process in plasma and to stabilize
neutron-excess nuclei.

Using  these parameters $(\langle {Z}\rangle=2, \langle
{A}/{Z}\rangle=5)$, we obtain that the solar core radius
${\mathbb{R}}\approx 9.4\cdot 10^{9}$ cm $({\mathbb{R}}/R_\odot
\approx 0.13)$ and its temperature ${\mathbb{T}}\approx 6.1\cdot
10^7 $ K. It is interesting that the solar core mass amounts to
$9.6\cdot 10^{32}~g$, i.e. almost exactly one half of full mass of
the Sun is concentrated in its core.

\section{The low frequency oscillation of the density of a neutral
plasma}

According to Eq.({\ref{neq}}) a hot plasma has the density
${\Theta}$ at its equilibrium state. The local deviations from
this  state induce processes of a density oscillation since the
plasma tends to return to its steady-state density. If we consider
small periodic oscillations of core radius
\begin{equation}
R={\mathbb{R}}+ u_R \cdot sin~ \omega_\Theta t,
\end{equation}
where a radial displacement of plasma particles is small
($u_R\ll{\mathbb{R}}$), the oscillation process of plasma density
can be described by the equation
\begin{equation}
\frac{d{\mathbb{F}}_{plasma}}{dR}={\mathbb{M}}\ddot R~.
\end{equation}
From this
\begin{equation}
\omega_\Theta^2=\frac{3\pi^{3/2}}{2}
k{\mathbb{T}}\biggl(\frac{e^2}{r_B k{\mathbb{T}}}\biggr)^{3/2}
\frac{r_B^3 {\Theta}}{{\mathbb{R}}^2\langle A/Z\rangle m_p}
\end{equation}
or
\begin{equation}
\omega_\Theta=\biggl\{\frac{\sqrt{\pi}~2^4}{\sqrt{10}~(3/2)^7}
\alpha^{3/2}\biggl[\frac{Gm_p}{r_B^3}\biggr]\biggl\langle\frac{A}{Z}\biggr\rangle\biggl[\langle
Z\rangle+1\biggr]^{4.5}\bigg\}^{1/2},\label{qm}
\end{equation}
where $\alpha=\frac{e^2}{\hbar c}$ is the fine structure constant.
These low frequency oscillations of neutral plasma density are
like to phonons in solid bodies. At that oscillations with
multiple frequencies $k\omega_\Theta$ can exist. Their power is
proportional to $1/k$, as the occupancy these levels in energy
spectrum must be reversely proportional to their energy
$k\hbar\omega_\Theta$. As result, low frequency oscillations of a
plasma density constitute set of vibrations
\begin{equation}
\sum_{k=1} \frac{1}{k}~sin(k\omega_\Theta t)~.
\end{equation}

\section{The main frequencies of the solar core oscillation}
The set of the low frequency oscillations with $\omega_\Theta$ can
be induced by sound oscillations with $\Omega_s$. At that
displacements obtain the spectrum:
\begin{equation}
u_R\sim \sin~\Omega_s t\cdot\sum_{k=0}
\frac{1}{k}~\sin~k\omega_\Theta t\cdot \sim \xi\sin~\Omega_s t +
\sum_{k=1} \frac{1}{k}~\sin~(\Omega_s \pm k \omega_\Theta
)t,\label{ur}
\end{equation}
where $\xi$ is a coefficient $\approx 1$.

This spectrum is shown in Fig.1b.

The central frequency of experimentally measured distribution of the solar
oscillation  is approximately equal to [Fig.1a]
\begin{equation}
{\cal F}_\odot\approx 3.23~ mHz\label{ffs}
\end{equation}
and the experimentally measured frequency splitting in this
spectrum is approximately equal to
\begin{equation}
{f}_\odot\approx 67.5~ \mu Hz\label{ffg}
\end{equation}
(see Fig.2b). At $\langle Z\rangle=2$ and $\langle A/Z\rangle=5$
the calculated frequencies of basic modes of oscillations (from
Eq.({\ref{qb}}) and Eq.({\ref{qm}})) are
\begin{equation}
{\cal F} =
\frac{\Omega_s}{2\pi}=3.19~mHz;~f_\Theta=\frac{\omega_\Theta}{2\pi}=66.0~\mu
Hz.
\end{equation}

It is important to note that there are two ways for the core
chemical composition determination. Thus chemical parameters
$\langle Z\rangle$ and $ \langle A/Z\rangle$ can be obtained by
fitting  at the known frequency of basic oscillation ${\cal F}$ as
it was done above (according to Table 1). Another way - to express
these parameters through ${\cal F}$ and $f$. The betweenness
relation of two this frequencies gives a possibility for a direct
determination of averaged parameters of nuclei  which the core
composed from:
\begin{equation}
\langle Z\rangle = \frac{4.49^{4/3}10^{1/3}10.5^{2/3}}{2^{8/3}\pi
\alpha}\biggl(\frac{f}{{\cal F}}\biggr)^{4/3}-1\label{z}
\end{equation}
and
\begin{equation}
\langle A/Z\rangle =\frac{3^7 \alpha^3 r_B^3}{5\cdot 10.5^3 G m_p}
\biggl(\frac{\pi^2 {\cal F}^3} {4.49^3 f^2}\biggr)^{2}\label{a/z}.
\end{equation}

Using  experimentally obtained frequencies (Eq.({\ref{ffs}}) and
Eq.({\ref{ffg}})), we have for the solar core
\begin{equation}
\langle Z\rangle_\odot = 2.03\label{zz}
\end{equation}
and
\begin{equation}
\langle A/Z\rangle_\odot = 4.99.\label{a/zz}
\end{equation}

Thus both these ways of determination of chemical parameters give
practically identical results that demonstrates the adequacy of
our consideration.
\bigskip

Author thanks  J.Christensen-Dalsgaard (Institut for Fysik og
Astronomi, Denmark) and G.Houdek (Institute of Astronomy of
Cambridge, UK) for help in  finding  publications of the
measurement data.

\begin{figure}
\begin{center}
\includegraphics[9cm,22cm][13cm,26cm]{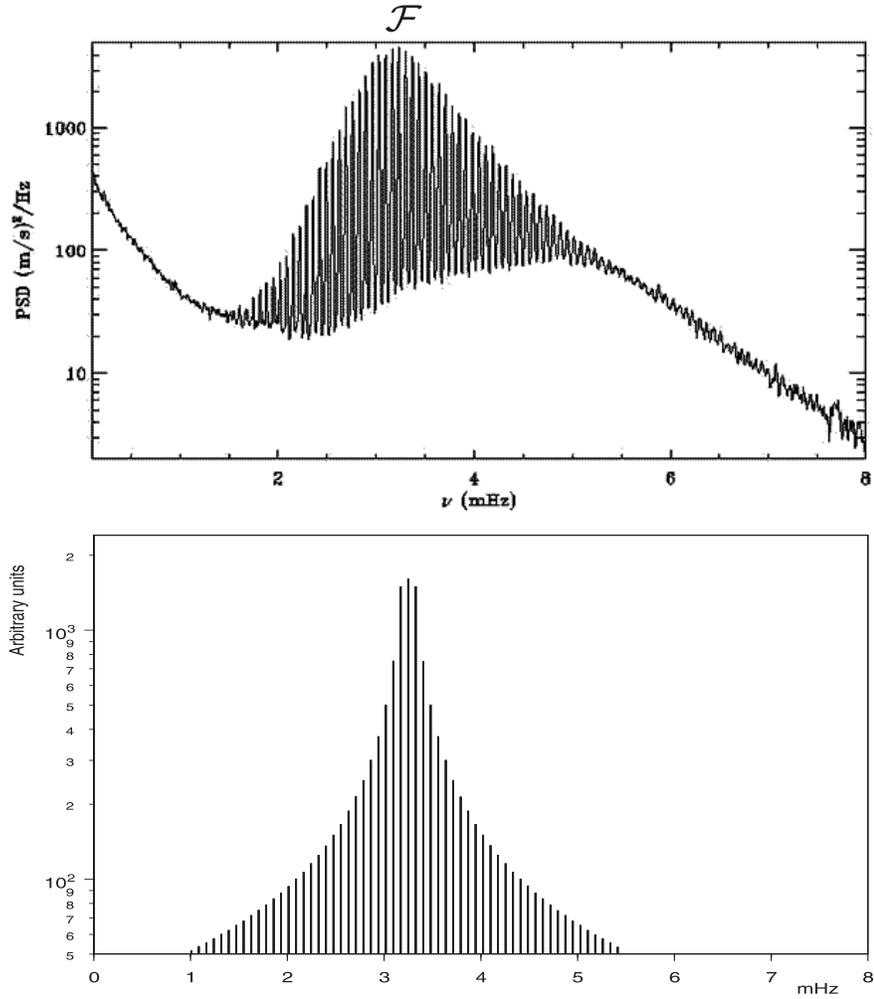}
\vspace{320 pt} \caption{ $(a)$ The measured power spectrum of
solar oscillation.  The data were obtained from the SOHO/GOLF
measurement \cite{5}. $(b)$ The calculated spectrum described by
Eq.({\ref{ur}}) at $Z=2$ and $A/Z=5$. }\label{spctr1}
\end{center}
\end{figure}

\clearpage

\begin{figure}
\begin{center}
\includegraphics[8cm,18cm][10cm,22cm]{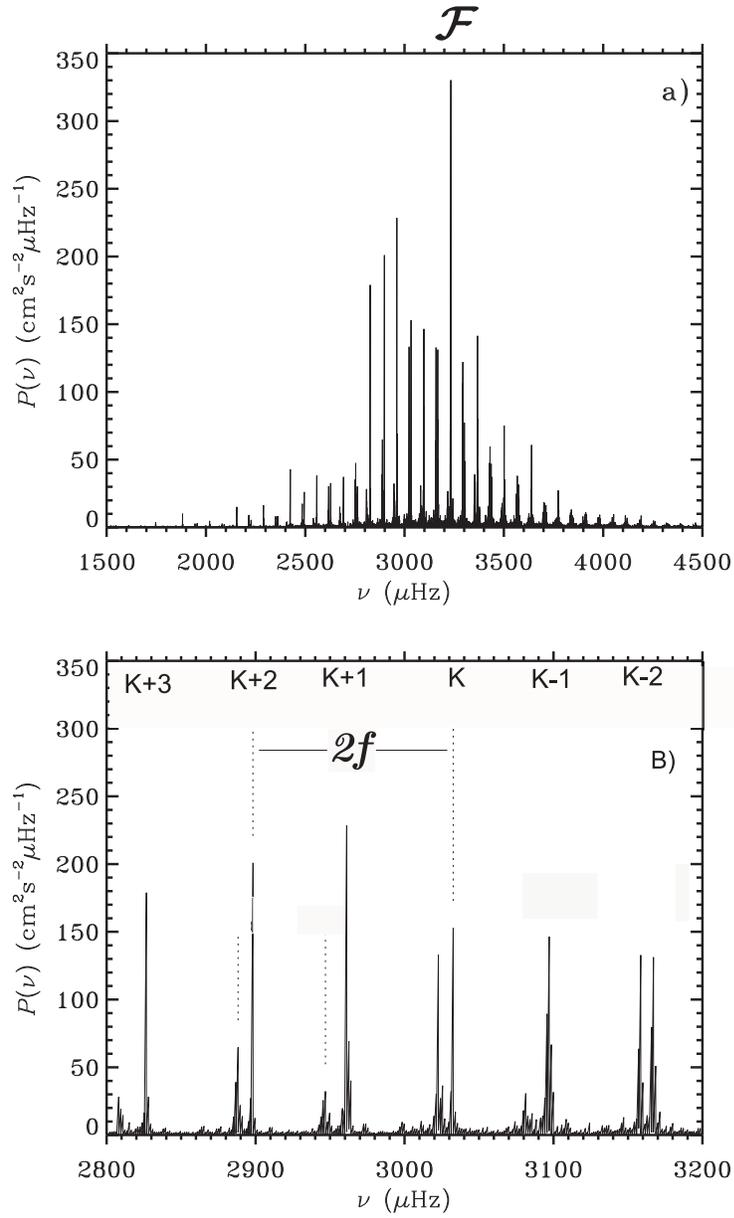}
\vspace{280 pt}\caption {$(a)$ The power spectrum of solar
oscillation obtained by means of Doppler velocity measurement in
light integrated over the solar disk. The data were obtained from
the BISON network \cite{6}. $(b)$ An expanded view of a part of
the frequency range.}\label{spctr2}
\end{center}
\end{figure}


\clearpage




\begin{thebibliography}{5}


\bibitem {1}    Vasiliev B.V. - Nuovo Cimento B, 2001, v.116, pp.617-634.
\bibitem {2}    Landau L.D. and Lifshits E.M. - Statistical Physics,1980, vol.II,3rd edition,Oxford:Pergamon.
\bibitem {3}    Landau L.D. and Lifshits E.M., - Hydrodynamics, vol. VI, (Addison-Wesley, Reading,Mass)
1965.
\bibitem {4}    Christensen-Dalsgaard, J., - Stellar oscillation, Institut for
Fysik og Astronomi, Aarhus Universitet, Denmark, 2003
\bibitem {5}    Solar Physics, vol.175/2, (http://sohowww.estec.esa.nl/gallery/GOLF)
\bibitem {6}    Elsworth, Y. at al. - In Proc. {\cal GONG'94} Helio- and Astero-seismology from Earth and Space,
eds. Ulrich,R.K., Rhodes Jr,E.J. and D\"{a}ppen,W., Asrtonomical
Society of the Pasific  Conference Series, vol.76, San
Fransisco,{\bf 76}, 51-54.

\end{thebibliography}
\end{document}